\documentclass[prl,twocolumn,nofootinbib, preprintnumbers, superscriptaddress]{revtex4}

\usepackage{amsmath,amssymb,slashed}
\usepackage{graphicx}
\usepackage{epstopdf}
\usepackage{float}
\usepackage[colorlinks=true,
            linkcolor=blue,
            urlcolor=blue,
            citecolor=green,          
            bookmarks=true,
            bookmarksnumbered=true,
            breaklinks=true,
            pdfpagemode=Fullscreen,
            pdfstartview=FitBH]{hyperref}

\usepackage[normalem]{ulem}
\usepackage{color}

\definecolor{Orange}{cmyk}{0,0.61,0.87,0}
\definecolor{JungleGreen}{cmyk}{0.99,0,0.52,0}
\definecolor{OliveGreen}{cmyk}{0.64,0,0.95,0.40}
\definecolor{Brown}{cmyk}{0,0.81,1,0.60}
\definecolor{RoyalBlue}{cmyk}{0.71,0.53,0,0.12}
\definecolor{Gray}{cmyk}{0,0,0,0.40}
\definecolor{LightPink}{cmyk}{0.0,0.25,0,0}
\definecolor{LLightPink}{cmyk}{0.0,0.10,0,0}
\definecolor{LightBlue}{cmyk}{0.25,0,0,0}
\definecolor{LightGray}{cmyk}{0,0,0,0.2}

\usepackage{xcolor}
\definecolor{gesfpurple}{rgb}{0.47,0.19,0.42}

\definecolor{gesflanse}{rgb}{0.00,0.50,0.50}

\definecolor{gesfblue}{rgb}{0.08,0.42,0.76}

\definecolor{gesfred}{rgb}{1,0,0}

\definecolor{gesfwhite}{rgb}{1,1,1}

\definecolor{gesfblack}{rgb}{0,0,0}

\newcommand{\geqn}[1]{Eq.\,\hypersetup{linkcolor=blue}(\ref{#1})\hypersetup{linkcolor=blue}}
\newcommand{\gfig}[1]{{\hypersetup{linkcolor=violet}Fig.\,\ref{#1}\hypersetup{linkcolor=blue}}}

\graphicspath{{figs/}}

\begin{document}

\title{Forbidden Dark Matter Combusted Around Supermassive Black Hole}

\author{Yu Cheng}
\email[Corresponding Author: ]{chengyu@sjtu.edu.cn}
\affiliation{Tsung-Dao Lee Institute \& School of Physics and Astronomy, Shanghai Jiao Tong University, China}
\affiliation{Key Laboratory for Particle Astrophysics and Cosmology (MOE) \& Shanghai Key Laboratory for Particle Physics and Cosmology, Shanghai Jiao Tong University, Shanghai 200240, China}

\author{Shao-Feng Ge}
\email[Corresponding Author: ]{gesf@sjtu.edu.cn}
\affiliation{Tsung-Dao Lee Institute \& School of Physics and Astronomy, Shanghai Jiao Tong University, China}
\affiliation{Key Laboratory for Particle Astrophysics and Cosmology (MOE) \& Shanghai Key Laboratory for Particle Physics and Cosmology, Shanghai Jiao Tong University, Shanghai 200240, China}

\author{Xiao-Gang He}
\email{hexg@sjtu.edu.cn}
\affiliation{Tsung-Dao Lee Institute \& School of Physics and Astronomy, Shanghai Jiao Tong University, China}
\affiliation{Key Laboratory for Particle Astrophysics and Cosmology (MOE) \& Shanghai Key Laboratory for Particle Physics and Cosmology, Shanghai Jiao Tong University, Shanghai 200240, China}
\affiliation{Department of Physics, National Taiwan University, Taipei 10617, Taiwan}

\author{Jie Sheng}
\email{shengjie04@sjtu.edu.cn}
\affiliation{Tsung-Dao Lee Institute \& School of Physics and Astronomy, Shanghai Jiao Tong University, China}
\affiliation{Key Laboratory for Particle Astrophysics and Cosmology (MOE) \& Shanghai Key Laboratory for Particle Physics and Cosmology, Shanghai Jiao Tong University, Shanghai 200240, China}

\begin{abstract}
The forbidden dark matter cannot annihilate into
a pair of heavier partners, either SM
particles or its partners in the dark sector,
at the late stage of cosmological evolution
by definition. We point out the possibility of reactivating the forbidden
annihilation channel around
supermassive black holes. Being attracted towards a black hole,
the forbidden dark matter is significantly accelerated to overcome
the annihilation threshold. The subsequent decay of the annihilation
products to photon leaves a unique signal around the black hole,
which can serve as a smoking gun for the forbidden dark matter.
For illustration, the Fermi-LAT data around Sgr $A^*$ provides
a preliminary constraint on the thermally averaged cross section
of the reactivated forbidden annihilation that is consistent
with the DM relic density requirement.
\end{abstract}

\maketitle 

{\bf Introduction} --
More than 80\% of the matter in our Universe today 
are dark matter (DM) \cite{Arbey:2021gdg}. But the nature of DM is still
a mystery and suggests new physics beyond the Standard
Model (SM) of particle physics \cite{Young:2016ala}. 
Various mechanisms
of DM production in the early Universe have been proposed.
The most classical one is the thermal freeze-out \cite{Lee:1977ua}, especially
for Weakly Interacting Massive Particle (WIMP).
In this scenario, DM with weak scale mass and interacting
strength can naturally predict the observed relic density,
which is called the WIMP Miracle.
There are also many other proposed DM 
production mechanisms to achieve the right relic density
\cite{Hall:2009bx,Garny:2017rxs,DAgnolo:2018wcn,Mizuta:1992qp,DEramo:2010keq,Belanger:2014bga,Bandyopadhyay:2022tsf,Cai:2015zza,Dey:2016qgf,Hashino:2021dvx,Bringmann:2021tjr,Xing:2021pkb,Puetter:2022ucx,Dodelson:1993je,Pospelov:2007mp,Hochberg:2014dra,Hochberg:2014kqa,Bernal:2017mqb,DAgnolo:2015nbz}. 
Of particular interest is the forbidden DM.

It was first noticed in \cite{Griest:1990kh} that 
the forbidden annihilation, where DM annihilates
into heavier particles, can happen due to the high-energy
tail of the thermalized velocity distribution in the
early Universe. With heavier annihilation product,
the forbidden channel is exponentially suppressed. 
This allows forbidden DM to have a stronger interaction
strength than the usual $\langle \sigma v \rangle \propto m_\chi$
scaling behavior of the WIMP freeze-out scenario.
A typical realistic model introduces a heavy dark photon
as a partner of the forbidden DM \cite{DAgnolo:2015ujb}.
In addition to the forbidden channel that determines the relic
abundance,  this model also naturally introduces DM 
self-interaction. With tiny mass splitting, either
positive or negative, this forbidden scenario has
a variation as impeded DM \cite{Kopp:2016yji}.
More concrete models can be found in
\cite{Delgado:2016umt,DAgnolo:2020mpt,Wojcik:2021xki},
in addition to an interesting $3 \rightarrow 2$ channel
\cite{Cline:2017tka}.

However, the DM Boltzmann distribution with typically
$\mathcal O(100)$\,km/s velocity dispersion in our 
galaxy nowadays can no longer support the forbidden
channel. Although it is possible to create other
annihilation channels into $e^+ e^-$ \cite{DAgnolo:2015ujb},
$\gamma \gamma$ \cite{Tulin:2012uq,Jackson:2013pjq,Jackson:2013tca},
and other heavier particles \cite{Jackson:2013pjq,Jackson:2013tca},
they are suppressed by kinematic mixing or loop factor.
The forbidden DM has a difficulty of being probed
via indirect detection.
One of the three major probes (direct, indirect, and
collider detections) of DM is almost completely missing. Then it
becomes very hard to verify whether the
DM, if found, is a forbidden type or not.
It is  of great importance to find a way
of testing the forbidden nature of such
scenario.

Around the supermassive black hole (SMBH), DM accretes and forms
a spike due to the strong gravitational potential.
This enhances the DM annihilation signals at the galaxy center (GC)
\cite{Gondolo:1999ef,Merritt:2003qk,Gnedin:2003rj,Regis:2008ij,Fields:2014pia,Sandick:2016zeg,Shelton:2015aqa,Johnson:2019hsm,Chiang:2019zjj,Yuan:2021mzi,Xia:2021tyf}. 
Meanwhile, the velocity of DM keeps increasing along its way towards the SMBH
\cite{Baushev:2008yz,Banados:2009pr}.
Comparison with the observed $\gamma$-ray flux from GC can
put a constraint on the DM annihilation cross section
\cite{Alvarez:2020fyo,HESS:2018cbt,HESS:2011zpk,Lefranc:2016srp}.
However, the forbidden annihilation has not been discussed yet.

We point out for the first time that SMBH
is a perfect place for testing the forbidden DM scenario. 
The SMBH acts as a natural particle accelerator 
and reactivates the forbidden annihilation into
heavier particles. With further decay into
photons, the DM spike then becomes a point source
of $\gamma$-ray.
We compare the predicted $\gamma$-ray spectrum
with the Fermi-LAT observation for the emission
around Sgr $A^*$ in our galaxy center to 
constrain the forbidden annihilation cross section.


{\bf Reactivating Forbidden Channel around SMBH} --
For illustration, we adopt
an economic scenario of forbidden DM with
one DM particle $\chi$, one heavy partner $F$, and
a light vector mediator $\phi_\mu$,
\begin{eqnarray}
  \mathcal L_{\mathrm{DM}} 
=
  g_{\chi \phi} \bar \chi \gamma^\mu \chi \phi_\mu
+ g_{F \phi} \bar F \gamma^\mu F \phi_\mu.
\label{eq:L}
\end{eqnarray}
The annihilation channel $\chi \bar \chi \rightarrow F \bar F$ is forbidden
if DM $\chi$ is ligher than the final-state $F$, $m_\chi \lesssim m_F$.
Typically, the DM velocity in galaxy is $10^{-3}$ of the
speed of light as determined by the galaxy gravitational
potential. Correspondingly, the DM kinetic energy is
only $10^{-6}$ of its mass. A relative mass difference
$\Delta_{F \chi} \equiv (m_F-m_{\chi})/m_{\chi}$
slightly larger than $10^{-6}$ can forbid DM annihilation
and the consequent indirect detection.

To reopen the forbidden channel, it is necessary to accelerate
the DM particles to overcome the annihilation threshold. This
naturally happens when DM gets accreted around black hole,
especially SMBH in the galaxy center.
A DM particle can be easily accelerated to become relativistic.
Especially, its velocity can reach the speed of light
around the event horizon \cite{Harada:2014vka,Banados:2009pr}.
With the help of a SMBH, forbidden DM can in principle overcome
the annihilation threshold no matter how large
the mass difference can be.

In its vicinity, the SMBH dominates gravitational potential.
Considering a DM particle attracted from infinity,
its velocity at distance $r$ from the SMBH
scales as $v(r) \propto r^{-1/2}$. Below we will give concrete
evaluation of the DM density and velocity profiles in detail.

\begin{figure}[t]
\centering
\includegraphics[width=0.486\textwidth]{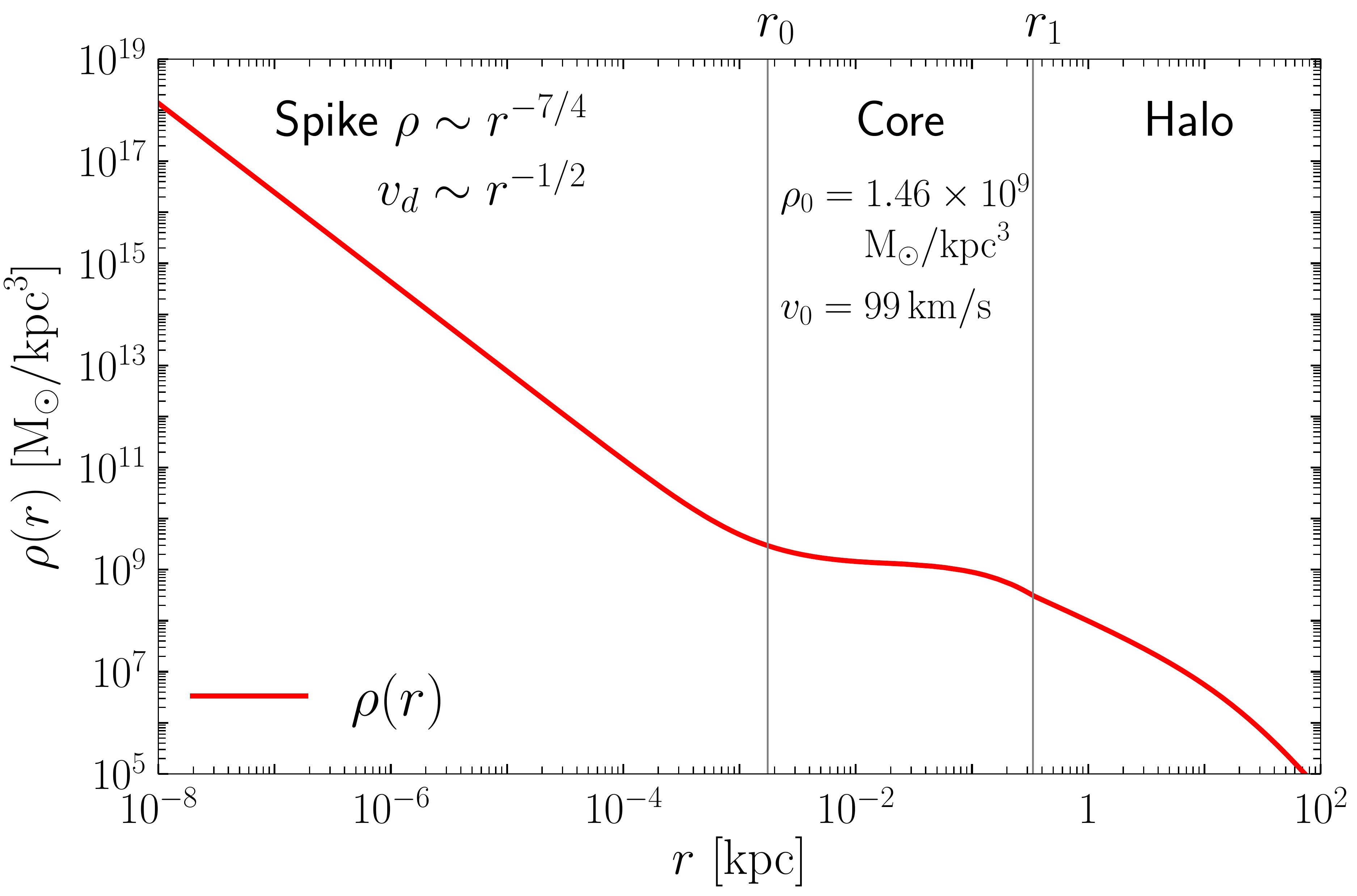}
\caption{The DM density profile $\rho(r)$ 
as a function of radius $r$. The grey vertical
lines divide the whole range into three regions: halo, core, and
spike.}
\label{fig:profile}
\end{figure}

With self-scattering, the DM velocity and density profiles have
more sophisticated dependence on the radius $r$.
We divide the whole galaxy into three regions,
\begin{equation}
  \rho(r)
=
\begin{cases}
  \mbox{ Halo:} \hspace{4mm} \rho_{\text {NFW }}(r), & r > r_1, \\
  \mbox{ Core:} \hspace{4mm} \rho_{\text {iso }}(r), & r_0 < r < r_1, \\
  \mbox{Spike:} \hspace{4mm} \rho_{\text{spike}}(r), & r_{in} < r < r_0,
\end{cases}
\end{equation}
as shown in \gfig{fig:profile}. 
The division $r_1$ is the characteristic radius
where the DM particle can experience one scattering on average
during the halo age $t_{\rm age}$ \cite{Kaplinghat:2015aga}. For $r > r_1$, the DM is treated
as collisionless and the Navarro-Frenk-White (NFW) profile
    \cite{Navarro:1995iw},
$\rho(r)= \rho_s (r / r_s)^{-1} (1+r / r_s)^{-2}$
with $\rho_s = 4.1 \times 10^6 M_{\odot}/$\,kpc$^3$
and $r_s = 26$\,kpc \cite{Alvarez:2020fyo,Abazajian:2020tww}, is a good approximation.

\textbf{Core:}
For the inner region, $r <r_1$, DM experiences frequent
collisions to maintain an isothermal core. Its DM density profile
is obtained by solving the Jeans equation \cite{Kaplinghat:2013xca,Kaplinghat:2015aga}.
On the boundary between the core and
halo regions, the density is the same as the NFW profile to make
the transition continuous. Since the NFW profile in the halo region
is kept as the original one, the mass enclosed within $r_1$ for the
isothermal solution should also remain the same as its NFW counterpart.
In other words, the NFW profile sets an initial condition for the
core region. At the center, the isothermal DM profile approaches a
constant density. With these conditions, the DM density
$\rho_0 \equiv \rho(r = 0)$ and the one-dimensional velocity dispersion
$v_{0}$ of the isothermal core can be determined.
In the current study, we take the transverse self-scattering
cross section \cite{Feng:2009hw,Tulin:2013teo,Schutz:2014nka,Tulin:2017ara} to be $\langle \sigma_T v \rangle /m_\chi
= 1.5 \,($cm$^2$/g$) \times $(km/s) for illustration. With a simple scaling $ \sigma_T \propto g_{\chi \phi}^4 / m_\chi^2$,  the coupling $g_{\chi \phi}$ increases with the DM mass $m_\chi$, $g_{\chi \phi} \propto m_\chi^{3/4}$. However, the coupling remains perturbative ($g_{\chi \phi} <1)$ in the whole range under consideration.
For the best-fit $m_\chi = 6.6\,$GeV from fitting the
Fermi-LAT data as we elaborate later, the corresponding
self-interaction coupling is $g_{\chi \phi} = 0.021$. With 10\,Gyr
galaxy age, the division between the halo and core regions happens at
$r_1 = 0.33\,$kpc.
The gravitational potentials of both baryons and dark matter are
taken into consideration \cite{Kaplinghat:2013xca}.
Then solving the Jeans equation gives a velocity dispersion 
$v_{0} = 99\,$km/s and density
$\rho_0 = 1.46 \times 10^9 M_{\odot}/$kpc$^3$
inside the core region.

\textbf{Spike:}
The DM core appears as a platform in the density profile.
However, the influence of SMBH leads to a DM spike in the
innermost region since the DM gas can no longer maintain
isothermal in such a strong and deep gravitational potential.
For $r < r_0 \equiv G M/v^2_{0}$, the gravitational
force of SMBH dominates. The DM density matches with the core
profile at $r_0$. Namely, the core profile provides an initial condition
for the spike region. The density profile $\rho(r)$ and the
one-dimensional velocity dispersion $v_d(r)$ inside the spike
can then be obtained by numerically solving the coupled differential
equations Eq.(33) and Eq.(34) of \cite{Shapiro:2014oha}. 
The exact solution follows power-law scaling
in the majority part of the spike, $\rho \propto r^{-(3+a)/4}$
and $v_d(r) \propto r^{-1/2}$. While the velocity scaling behavior
is consistent with our naive expectation earlier and hence
can apply universally, the density scaling is related
to the velocity dependence of the self-scattering
cross section $\sigma = \sigma_0 (v_d / v_{0})^a$ via $a$.
For illustration, a light mediator $\phi$ with mass
$m_\phi \sim 10^{-3} m_\chi$ predicts $a = 4$ and correspondingly 
$\rho \propto r^{-7/4}$. \gfig{fig:profile} shows our
numerical evaluation of the spike DM density profile including
relativistic effects. However, the DM density cannot
increase forever towards the SMBH. It has a sharp cutoff
and vanishes at an inner boundary $r_{in} \equiv 4GM$
\cite{Sadeghian:2013laa,Shapiro:2014oha}.

From the core platform to the vicinity around SMBH, the DM
density increases by at least 9 orders as shown in
\gfig{fig:profile}. Correspondingly, the DM self-scattering
rate scales as $\rho^2(r)$ and increases by at least 18 orders.
So the DM fluid around SMBH is fully thermalized.
Being relativistic, the DM velocity follows
the J{\"u}ttner distribution \cite{Juttner:1911,DeGroot:1980dk},
\begin{equation}
  f_J({\bf p})
=
  \frac 1 {4 \pi T_\chi m^2_\chi K_2(x)}
  e^{- \frac {\sqrt{|{\bf p}|^2 + m^2_\chi}}{T_\chi}},
\label{eq:fJ}
\end{equation}
where the temperature parameter $x \equiv m_\chi / T_\chi$ is the ratio of DM mass $m_\chi$
and temperature $T_\chi$.
The DM temperature is equivalent to its velocity dispersion,
$T_\chi = \frac 3 2 m_\chi v^2_d$.
With $v_d(r) \propto r^{-1/2}$, the DM temperature
($T_\chi \propto 1/r$) keeps increasing towards the SMBH
to overcome the forbidden annihilation threshold.

{\bf The Forbidden DM Signature} --
With the forbidden channel $\chi \bar \chi \rightarrow F \bar F$
being reactivated, the final-state $F$ further decays into
neutrino and photon ($F \rightarrow  \nu \gamma$) through the operator $g_{F \nu A} \bar{F} \sigma^{\mu \nu} \nu \mathcal{F}_{\mu \nu}$ where $\mathcal{F}_{\mu \nu}$ is the photon field strength.
The photon can be collected by astrophysical $\gamma$-ray
observations at Fermi-LAT. 
In principle, the DM particle 
$\chi$ can also annihilate into a 
pair of mediators, $\chi \bar 
\chi \rightarrow \phi \phi$. At 
first glance, the light mediator 
can form ladder diagrams to 
enhance the annihilation cross 
section by the Sommerfeld 
enhancement \cite{Hisano:2002fk,Hisano:2003ec,Hisano:2006nn,Arkani-Hamed:2008hhe}. However, 
this enhancement only happens 
with non-relativistic DM, $v^2 
\leq \alpha_d m_\phi/ m_\chi$ 
with $\alpha_d \equiv g_{\chi 
\phi}^2/(4 \pi)$, and hence does 
not apply here. In addition, this 
channel has no observable signal 
as will be detailed later. Unless 
stated otherwise, we focus on the 
forbidden channel $\chi \bar \chi \rightarrow F \bar F$.

With DM particles being fully thermalized and following the Juttner distribution
in \geqn{eq:fJ}, the kinematics of a DM pair with momentums
$p_1$ and $p_2$ can be parameterized
in terms of the total invariant mass squared $s \equiv (p_1 + p_2)^2$ or
equivalently the relative velocity $V_r$, the centre-of-mass
velocity $V_c$ 
or equivalently the energy sum $E_+ \equiv E_1 + E_2$, and
their energy difference $E_- \equiv E_1 - E_2$. The other three degrees
of freedom in ${\bf p}_1$ and ${\bf p}_2$ can be reduced by
symmetry considerations: 1) the overall two directional angles
of ${\bf p}_1 + {\bf p}_2$ by the spherical invariance of
the whole system, and 2) another azimuthal angle by the
rotational invariance around the total moment. 

It is more convenient to first calculate the differential
cross section in the center-of-mass frame where both the
initial $\chi$ and the final $F$ have fixed energy,
$\sqrt s / 2$. For the $s$-channel annihilation dictated by
\geqn{eq:L},
\begin{eqnarray}
  \frac{d \sigma}{d \Omega}
=
  g_{\chi \phi}^2 g_{F \phi}^2
  \sqrt{\frac {s_F}{s_\chi}}
  \frac{s(4 m_F^2 + 4 m_\chi^2 +s ) + \cos^2 \theta^c_F s_F s_\chi}
  {64 \pi^2 s (s - m_\phi^2)^2},
\end{eqnarray}
with $s_F \equiv s - 4 m_F^2$ and $s_\chi \equiv s - 4 m_\chi^2$.
The angular dependence part $\cos \theta^c_F$,
where $\theta^c_F$
is the scattering angle between the $F$ and total momenta,
is highly suppressed by $s_F$ and $s_\chi$. For $m_\chi = 10$\,GeV,
$m_F = 11$\,GeV, and a typical $s = 1.3 \times 4 m_F^2$, the angular
dependent part only contributes 2\%. For simplicity, we treat
this process as isotropic.

The produced $F$ further decays into photons with fixed
energy $m_F / 2$ in the $F$ rest frame. Lorentz boost
back to the center-of-mass frame renders a box-shaped
spectrum\cite{Ibarra:2012dw,Ibarra:2013eda}, $d N_\gamma/ d E_\gamma = 1 / (E^c_+ - E^c_-)$
where limits $E^c_\pm \equiv m_F \gamma_F (1 \pm v_F) / 2$ with
$\gamma_F \equiv 1 / \sqrt{1 - v^2_F}$ and
$v_F \equiv \sqrt{1 - 4 m^2_F / s}$. Since the primary
process $\chi \bar \chi \rightarrow F \bar F$ is
isotropic in the center-of-mass frame, the box-shaped
photon spectrum is also isotropic.

It is necessary to further Lorentz boost to the lab frame
where the photon spectrum is observed. Due to isotropy,
photon of any given energy $E^c_\gamma$ in the center-of-mass frame
again possesses box-shaped spectrum in the lab frame,
$d N_\gamma / d E_\gamma = 1 / (E^L_+ - E^L_-)$ with limits
$E^L_\pm \equiv E^c_\gamma \gamma_c (1 \pm v_c)$.
The final photon spectrum in the lab frame is then a convolution
of these two-step box spectra.
\begin{figure}[!t]
\centering
\includegraphics[width=.47\textwidth]{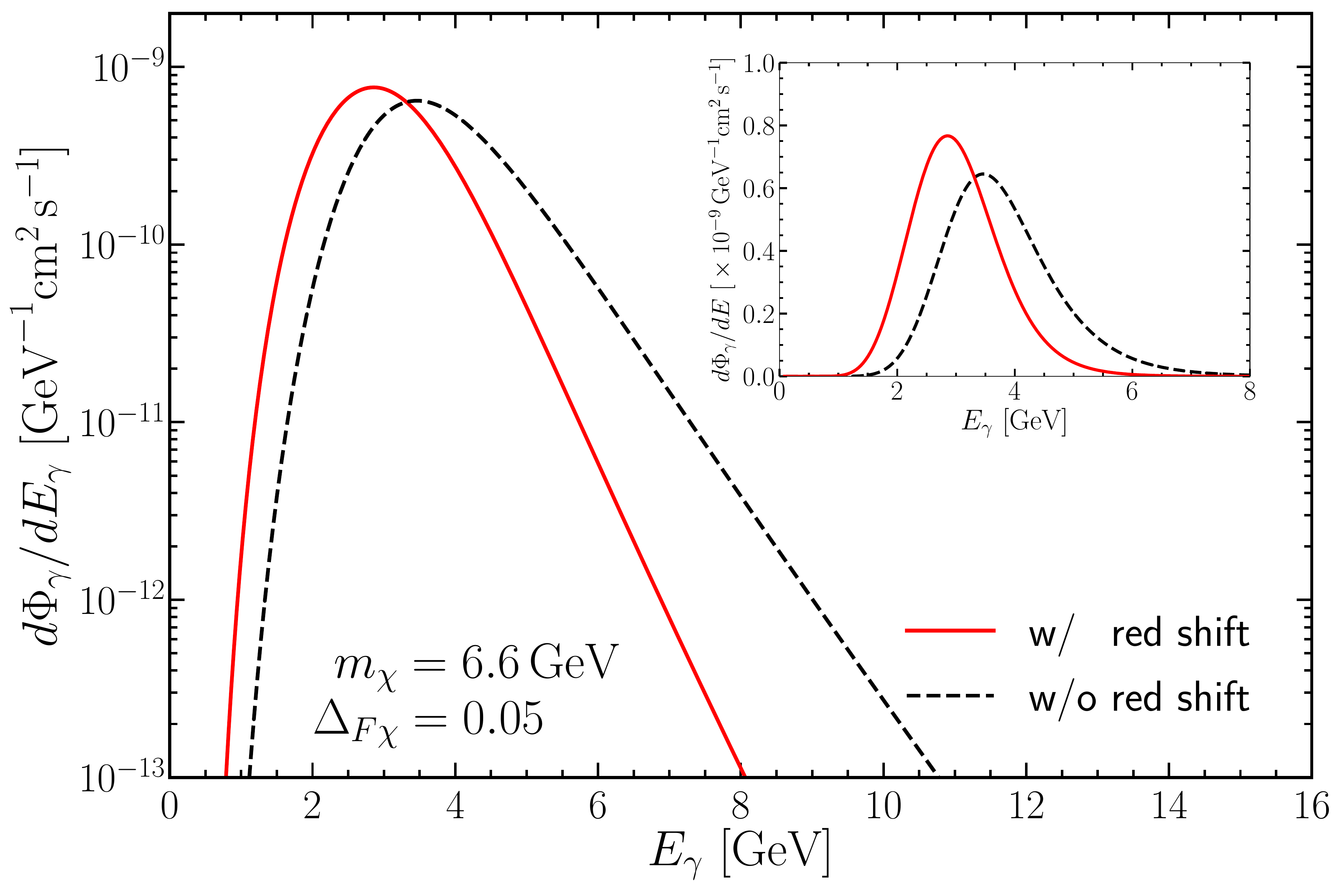}
\caption{The $\gamma$-ray spectrum $d \Phi_\gamma / d E_\gamma$
illustrated with DM mass $m_\chi = 6.6$\,GeV and the relative
mass difference $\Delta_{F \chi} = 0.05$.
Correspondingly, the total flux injected is
$1.4 \times 10^{-9}$\,cm$^2$/s which
is the best-fit value of the Fermi-LAT data.
Both results with or without redshift are shown as
red solid and black dashed lines, respectively. The insert plot
shows the same results in linear scale.
}
\label{fig:Energy_Spectrum}
\end{figure}

Since the two initial DM particles are fully thermalized,
thermal average is necessary to obtain the photon spectrum, 
\begin{equation}
  \frac{d F_\gamma}{d E_\gamma} (r)
=
  \int_0^1 d V_{r} d V_{c}
  P_{r}
  \left(V_{r},V_c,x\right)
  \sigma V_{r}
  \frac{dN_{\gamma}}{dE_{\gamma}}(V_{r},V_c).
\label{eq:relativistic_flux}
\end{equation}
%
%
For evaluating the total flux from DM annihilation, one needs
to integrate over the radius $r$. However, not all DM
particles can overcome the annihilation threshold.
At a large radius, only the high-energy tail of the
Juttner distribution can contribute. So there is no need
to integrate the radius $r$ to infinity. We may
set an upper limit $r_b$ on the radius integration.
The concrete value of $r_b$ can be numerically solved
by requiring the local phase space to have at least 1\%
contribution above the annihilation threshold,
\begin{equation}
  \int_0^{V^{\rm th}_r} d V_r \int_0^1 d V_c P_r(V_r, V_c, x(r_b))
\equiv
  0.99,
\label{eq:defrb}
\end{equation}
where
$V^{\rm th}_r \equiv [1- 1/ (1-2 m_F^2/m_\chi^2)^2]^{1/2}$
is the relative velocity at threshold. 
Note that the temperature
parameter $x(r)$, as defined below \geqn{eq:fJ}, is a function of the radius since $T_\chi$ is a function of $r$.
The joint probability distribution,
\begin{equation}
  P_{r}
  \left(V_{r}, V_c, x\right)
\equiv
 \frac{x^2}{K_2^2(x)}
  \frac{\gamma_{r}^3 
  (\gamma_{r}^2-1) V^2_c}{(1-V_c^2)^{2}}\,
  e^{-x \sqrt{\frac {2  + 2 \gamma_r}{1-V_c^2}}},
\end{equation}
reduced from the two DM Juttner distributions
with momenta $p_1$ and $p_2$, is a function of
their relative velocity $V_r \equiv \sqrt{\gamma^2_r - 1} / \gamma_r$
($\gamma_r \equiv (p_1 \cdot p_2) / m^2_\chi$)
and center-of-mass velocity 
$V_c \equiv \sqrt{1 - s / (E_1 + E_2)^2}$ 
\cite{Cannoni:2013bza,Cannoni:2015wba}.
For those regions with $r > r_b$ and hence less than
$1\%$ of the total phase space that can contribute,
we omit their contributions for simplicity which is a
good enough approximation. While the integration volume
increases as $r^3$, the DM mass density factor
$\rho^2(r) \sim r^{-7/2}$ decreases even faster as we demonstrate below.
Consequently, the actual precision should be better
than 1\%.

The photon flux observed at Earth is then an integration,
\begin{equation}
  \frac{d \Phi_{\gamma}}{d E_{\gamma}}
=
  \frac{1}{4 \pi D^2} \frac{1}{4 m_\chi^2}
  \int^{r_b}_{r_{in}} 4 \pi r^{2}  d r
  \rho^2(r)
  \frac{dF_{\gamma}}{dE_{\gamma}}(r),
\label{eq:gamma_flux}
\end{equation}
with dilution by the distance $D$ to the SMBH.

Due to the extremely strong gravitational potential around
the SMBH, the emitted photon will lose energy when
traveling towards us. This gravitational redshift
reduces the photon energy by a factor of 
$\sqrt{1-2 G M / r_e}$ as a function of the radius
$r_e$ where the photon emitted. 
\gfig{fig:Energy_Spectrum} shows 
the photon fluxes in \geqn{eq:gamma_flux} 
with (red solid) and without (black dashed) redshift.
As expected, redshift drives the flux toward lower energy
with a narrower width. It is interesting to
see that the signal is actually a peak. Such a prominent
feature can help itself to be identified in observations
as we elaborate below.

\begin{figure}[!t]
\centering
\includegraphics[width=0.486\textwidth]{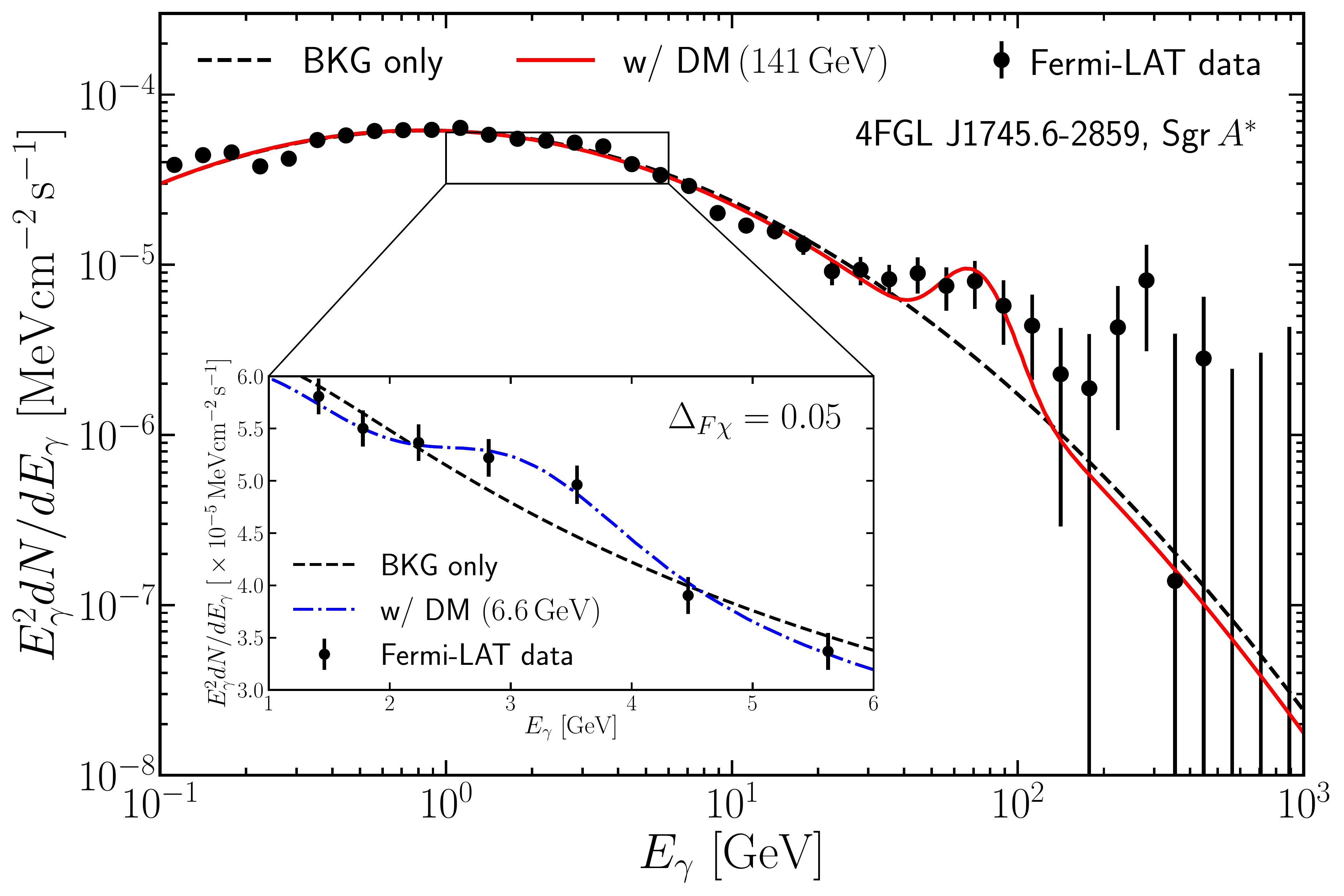}
\caption{
The Fermi-LAT data (black points with error bar),
background-only fit (black dashed line), and the signal fit 
with forbidden DM annihilation for
$m_\chi = 141\,$GeV (red solid) and $m_\chi = 6.6\,$GeV (blue dash-dotted).
For illustration, the mass difference between DM and its partner is
taken to be $\Delta_{F \chi} = 0.05$. 
}
\label{fig:fermiData}
\end{figure}

{\bf Constraints from FermiLAT} --
The predicted photon flux is localized around the SMBH.
We use the Fermi-LAT data to illustrate its constraint
on the Forbidden DM.
Being an all-sky $\gamma$-ray telescope,
Fermi-LAT has very good energy and angular resolutions on the
$\gamma$-ray point source from the GC \cite{Fermi-LAT:2009ihh,Fermi-LAT:2016uux}.
We analyze a total region of $10^\circ \times 10^\circ$ square
centered around the direction of Sgr\,$A^{*}$ and find several
$\gamma$-ray point sources, of which 4FGL\,J1745.6-2859 is the
brightest and closest one to Sgr\,$A^*$ in the Fourth catalog
of Fermi-LAT sources (4FGL) \cite{Fermi-LAT:2019yla,Ballet:2020hze}.
Also, this point source
is considered as the manifestation of Sgr\,$A^*$ in the
MeV-to-GeV range \cite{Cafardo:2021pqs}. We use 14
years of Fermi-LAT data from August 4, 2008 to October 26, 2022.
To be specific, the Pass 8 SOURCE-class events from 100\,MeV
to 1000\,GeV are binned to a pixel size of $0.08^\circ$.

A universal model can describe the $\gamma$-ray spectrum
of different point sources. The 4FGL\,J1745.6-2859 spectral
model is a log-parabola in the 4FGL Catalog
\cite{Fermi-LAT:2019yla,Ballet:2020hze},
\begin{equation}
  \frac{d N}{d E}
=
  N_0 \left( \frac E {E_{0}} \right)^{-\alpha-\beta \log (E / E_{0})},
\end{equation}
where $N_0$ is normalization, $E_0$ a scale parameter,
$\alpha$ the spectral slope at $E_0$, and $\beta$ the
curvature of the spectrum. Since $E_0$ does not vary
much, its value is fixed to 4074\,MeV in our fit while
the other three can freely adjust.
\gfig{fig:fermiData} shows the observed data of energy
spectrum from Fermi-LAT and the best-fit line using
the universal model. The black dashed line shows that
the background-only hypothesis can fit the data
pretty well with $\chi^2_{\rm min} = 140.8$.

The strength of the forbidden channel can be parametrized
in terms of the {\it weighted thermally 
averaged cross section} $\overline{\langle \sigma v \rangle}$,
\begin{equation}
  \overline{\langle \sigma v \rangle} 
\equiv 
  \frac{\int^{r_b}_{4GM} 
  4 \pi r^2 \rho^2(r) 
  \langle\sigma v (r) \rangle}
  {\int^{r_b}_{4GM} 4 \pi r^2 \rho^2(r)},
\end{equation} 
by taking the DM density $\rho(r)$ and geometrical measure
into consideration. Using $\chi^2$ minimization, 
we find two local best-fit points
at $m_\chi = 6.6$\,GeV and $\overline{\langle \sigma v \rangle} = 2.56 \times 
10^{-25}$\,cm$^{3}$\,s$^{-1}$
as well as $m_\chi = 141$\,GeV and $\overline{\langle \sigma v \rangle} = 5.32 
\times 10^{-24}\,$cm$^{3}$\,s$^{-1}$.  For the global fit point at $m_\chi = 6.6$\,GeV, the coupling constants are $g_{\chi \phi} = 0.021$ and $g_{F \phi} = 1.01$.
Compared with the background-only hypothesis, 
$\chi^2_{\rm min}$ decreases by $17.0$ and $14.2$, respectively.
\gfig{fig:fermiData}
shows the best-fit energy spectrum with DM hypothesis
as blue dash-dotted and red solid lines. We can clearly
see the prominent bump
around $E_\gamma \approx 3$\,GeV which corresponds to
the $m_\chi = 6.6$\,GeV best-fit value.
Although the $E_\gamma = 60$\,GeV peak in the observed
spectrum seems more prominent in \gfig{fig:fermiData}, it
actually has lower significance due to larger uncertainty.

\begin{figure}[!t]
\centering
\includegraphics[width=0.486\textwidth]{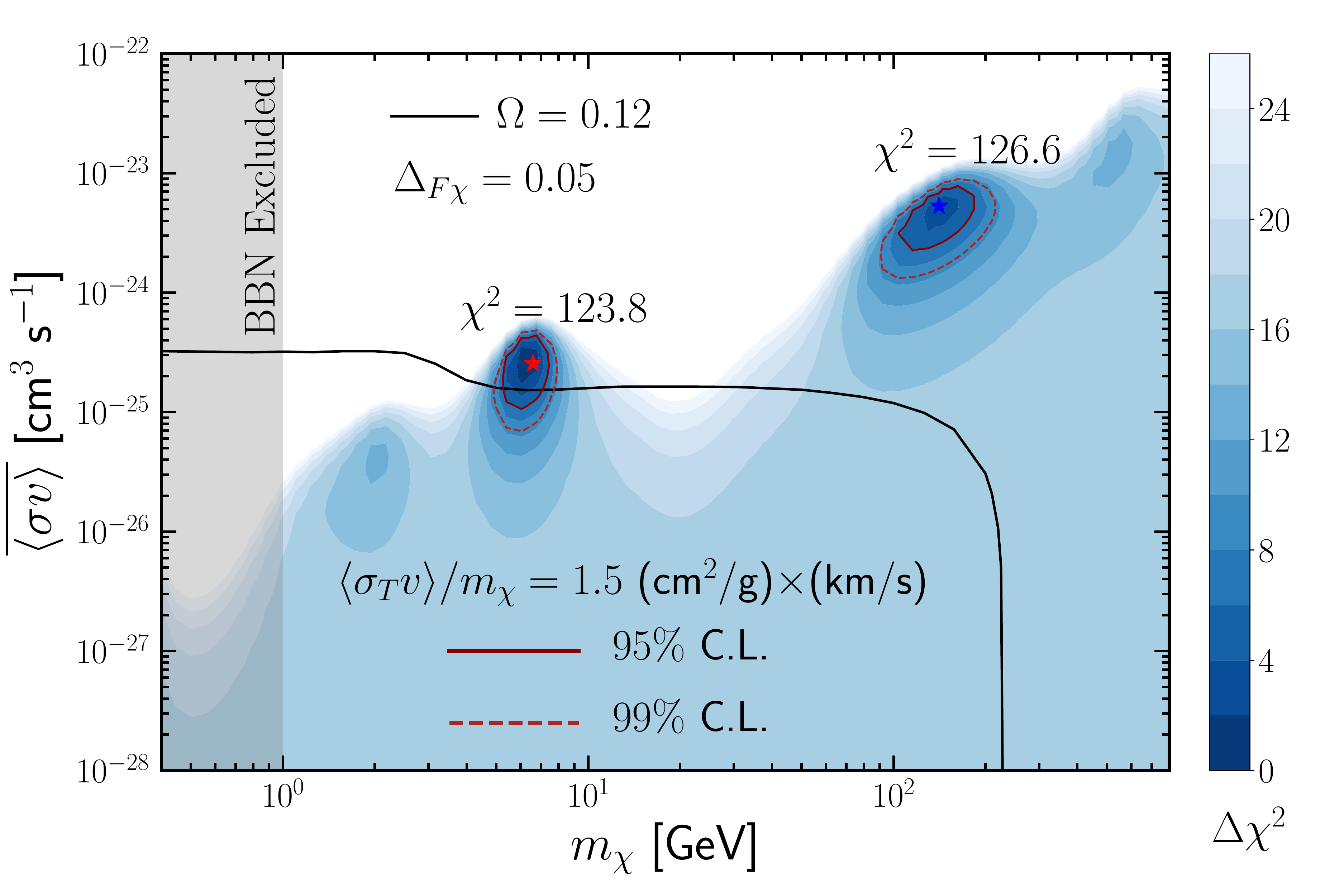}
\caption{ 
The sensitivity contours of forbidden DM annihilation 
cross section obtained by fitting the 
Fermi-LAT data with mass difference 
$\Delta_{F\chi} = 0.05$, $m_\phi = 10^{-3} m_\chi$, 
and self-scattering
cross section $\langle \sigma_T v \rangle / m_\chi
= 1.5$\,(cm$^2$/g)$\times$(km/s).
The red star represents the best-fit
point where $m_\chi = 6.6\,$GeV and 
$\overline{\langle \sigma v \rangle} = 2.56 \times 
10^{-25}$\,cm$^{3}$\,s$^{-1}$. 
The red solid and dashed lines are the sensitivity contours at
$95\%$ and $99\%$ C.L., respectively. 
The black solid line corresponds to the 
parameters that generate the correct relic density
$\Omega_{\rm CDM} = 0.12$ through freeze-out.
}
\label{fig:Limit_Contour}
\end{figure}

\gfig{fig:Limit_Contour} shows the signal sensitivities
obtained with $\Delta_{F\chi} = 0.05$ and   
self-scattering cross section
$\langle \sigma_T v \rangle / m_\chi
= 1.5\,$cm$^2$/g$\,\times\,$km/s. The sensitivity around
the best-fit point with $m_\chi = 6.6$\,GeV is quite
significant. We show the 95\% (red solid) and 99\%
(red dashed) contours to make it transparent. With a peak
structure in the inset of \gfig{fig:Energy_Spectrum},
the forbidden DM scenario can nicely explain the local
excess around $E_\gamma \approx 3$\,GeV.

{\bf Relic Density} --
The dark sector can be in thermal bath through the decay and inverse decay of $F$ to neutrino and photon. 
This requires the decay rate of $F$, $\Gamma \sim g^2_{F \nu A} m_F^3/ (8 \pi)$ to be larger than the Hubble constant, $H \propto T^2/M_{\rm pl}$ where $M_{\rm pl}$ is the Planck mass. By taking the freeze-out temperature $T \sim m_\chi / 25 \sim m_F/25$, we need 
$g_{F \nu A} \gtrsim 0.2 / \sqrt{M_{\rm pl} m_F} \sim (10^{-12}\sim 10^{-10})\,$GeV$^{-1}$ as a function of $m_F$. Since $g_{F \nu A}$ is a free parameter with no additional constraint, the requirement can be easily satisfied.
Further, the DM particle $\chi$ can also be in  thermal equilibrium via the annihilation processes $\chi \bar{\chi} \leftrightarrow F \bar F$. For simplicity, we consider symmetric DM with vanishing chemical potential.

With the same mass difference $\Delta_{F\chi}$ and
self-scattering cross section
$\langle \sigma_T v \rangle$,
the allowed annihilation strength
 $\overline{\langle \sigma v \rangle}$ and DM mass $m_\chi$ for
generating the correct relic density $\Omega_{\rm CDM} = 0.12$ through freeze-out 
is shown as black line in \gfig{fig:Limit_Contour}.
Both $\chi \bar{\chi} \rightarrow F \bar{F}$ and  $\chi \bar{\chi} \rightarrow \phi \phi$ can happen in the early Universe to determine the DM relic density. With fixed DM self-scattering cross section $\langle \sigma_T v \rangle / m_\chi
= 1.5$\,(cm$^2$/g)$\times$(km/s), the coupling $g_{\chi \phi}$ between $\chi$ and $\phi$ increases with the DM mass $m_\chi$.
Being modulated by the same coupling, the cross section of $\chi \bar{\chi} \rightarrow \phi \phi$ also increases with $m_\chi$. In other words, the DM freeze-out is dominated by the second annihilation channel for heavy DM and the first channel should have decreasing cross section in order to obtain the correct relic density. At some point, no parameter space is left for the first channel which means the involved coupling $g_{F \phi}$ should vanish. This explains why the black curve suddenly drops to zero.

We can see that the constraint 
from FermiLAT is consistent with the one from 
the DM relic density for $m_\chi = 6.6\,$GeV. However,
this is not claiming that the forbidden DM has already
been established by Fermi-LAT. More realistic analysis
by the experimental collaboration is necessary which
should give much more stringent bounds on
$\overline{\langle \sigma v \rangle}$.


The massive mediator $\phi$ can decay into two neutrinos
through loop correction with the massive fermion $F$ in the triangle diagram. 
Taking the typical values $g_{F \nu A} = 10^{-3}\,$GeV$^{-1}$ and $g_{F \phi} = 10^{-3}$, the predicted rate 
$\Gamma_{\phi} \simeq g_{F \phi}^2 g_{F \nu A}^4 m_F^4 m_\phi /48 \pi^3$ indicates that
the mediator can decay 
within $10^{-5}$\,s
and 
hence does not contribute to the relic density. 
However, a light mediator with mass $m_\phi \lesssim 1$\,MeV can cause extra relativistic degrees of freedom.
The grey band in \gfig{fig:Limit_Contour} shows 
the parameter region constrained by BBN.
Although the two-photon final state is also kinematically allowed, it is forbidden by the Landau-Yang theorem \cite{Landau:1948kw,Yang:1950rg} since $\phi$ is a vector boson.

{\bf Conclusion and Discussion} -- 
We propose using SMBH as a natural accelerator to
reactivate the forbidden DM to allow them to annihilate
into heavier partners. The subsequent decay into
$\gamma$ photons can leave unique signature for
astrophysical observation, such as those at Fermi-LAT.
Two important features can be
identified. First, the photon energy is mainly
determined by the DM mass and its spectrum has a
peak.
Second, such a signal is always associated with
SMBH instead of random distribution. If the angular
resolution is good enough, more geometrical features
can be identified.

\section*{Acknowledgements}

The authors would like to thank Xun Chen, Yu-Chen Wang,
and Hai-Bo Yu for useful discussions.
This work is supported by the 
National Natural Science Foundation of China (12375101, 12375088, 12090060, and 12090064), the Shanghai Pujiang Program (20PJ1407800), and 
the Double First Class start-up fund
(WF220442604). XGH was also supported in part by the MOST
(Grant No. MOST 106- 2112-M-002- 003-MY3 ). 
SFG is also an affiliate member of Kavli IPMU, University of Tokyo.

\vspace{15mm}
\end{document}